\documentstyle[11pt,newpasp,twoside]{article}
\def\edcomment#1{\iffalse\marginpar{\raggedright\sl#1\/}\else\relax\fi}
\reversemarginpar

\begin{document}
\title{HI 21cm-line observations with the GMRT towards interstellar clouds previously seen in optical absorption }
\author{Rekhesh~Mohan$^1$, K.S. Dwarakanath$^1$, G. Srinivasan$^1$ and Jayaram N. Chengalur$^2$}
\affil{$^1$Raman Research Institute, Bangalore, India\\
$^2$National~Center~for~Radio~Astrophysics, Pune, India}

\begin{abstract}
We have made HI
21cm-line absorption measurements using the GMRT towards 15 directions
in the Galaxy which are known to have high random velocity clouds
as seen in the optical absorption lines of CaII and NaI.
For the first time, in 6 out of these 15 directions we detect HI absorption
features corresponding to the high random velocity optical absorption
lines. The mean optical depth of these detections is $\sim$ 0.08.
\end{abstract}

\section{Introduction}
Our picture of the interstellar medium, its structure and motion emerged from the optical 
absorption studies in the lines of CaII and NaI towards bright O \& B stars. 
Soon 
after the 21cm-line from atomic hydrogen was discovered, attempts were made to 
detect the HI line from these clouds through emission measurements. 
Interestingly, although the HI emission velocity is clearly correlated with 
that of the optical absorption line at low random velocities, no HI was 
detected corresponding to the higher random velocity ($|v|$ $\ge$ 10 km sec$^{-1}$) 
optical absorption lines. Recently, an HI
absorption study was carried out (Rajagopal et al 1998a) using the VLA
towards radio sources in the sky which are aligned very close ($\sim$ a few arc 
mins) to the star. 
No HI absorption corresponding to the high random velocity optical
absorption lines was detected down to an optical depth of 0.1. 

\section{Motivation for the present observations}
To explain the absence of the HI absorption line corresponding to the higher 
random velocity optical absorption features, Rajagopal et al (1998b) advanced 
the following arguements based on the scenario suggested by Radhakrishnan \& 
Srinivasan (1980). If the high velocity optical absorption features arise in the
clouds which had been shocked and accelerated by supernova remnants in their 
late phases of evolution, then it would result in the higher random velocity 
clouds being warmer and also of lesser HI column density as compared to the low
random velocity clouds (shock heating and evaporation). The combination of the 
low column density and the higher temperature makes it tough to detect them in 
HI absorption. Our observations with the Giant meterwave radio telescope (GMRT) 
were primarily intended as a deeper HI absorption search, with a view to shed 
light on the above hypothesis. 

\section{Source selection and Observations}
We selected 15 directions, towards which both low and high velocity 
optical absorption lines were seen (Adams 1949, Welty et al 1994, 1996)
and towards which a bright radio source ($S_{21cm}$ $>$ 100 mJy) was aligned 
very close to the star.
Rajagopal et al (1998a) chose their directions such that at half the distance 
to the star, the linear separation between the lines of sight towards the star 
and the radio source was $\sim$ 3 pc. In the present observations, we have 
chosen the directions where this value is $<$ 1 pc in most cases. This gives a 
better chance for both the lines of sight to sample the same gas. We aimed at 
an rms in optical depth of about 0.01, about 10 times better than the previous
limits. The observations were carried out using 10 antennas of the 14 which are
located within the central 1 km${^2}$ of the GMRT. We used a bandwidth of 2 MHz
and 128 channels. Frequency switching on a nearby calibrator was employed for 
bandpass calibration. For each of the directions, the integration time varied 
from $\sim$ 1 hour to $\sim$ 7 hours, depending on the strength of the 
background source. Continuum subtracted spectral cubes were made for each of the
observed fields and the rms noise levels in the line images were found to be 
consistent with the expected values.

\section{Preliminary Results}
We have obtained the first detections of HI absorption features corresponding 
to the high random velocity optical absorption lines, in 6 out of the 15 
directions that we observed. 
The mean optical depth of these detections is $\sim$ 0.08. This low value of HI
optical depth is not typical for the well known diffuse interstellar clouds.\\
\\
The true nature of the high random velocity clouds identified through the 
optical absorption lines is still unclear. As a part of the ongoing studies, 
we will combine the HI absorption results with HI emission data and try to get 
a handle on the spin temperature and HI column density of these clouds. 
This would help us to address the question about the existence of a
population of shocked clouds.

\end{document}